\documentclass[12pt]{article}
\usepackage[round]{natbib}
 
\usepackage{latexsym}
\usepackage{graphicx}
\usepackage{amsmath}
\usepackage{amsfonts}
\usepackage{amssymb}
\usepackage{mathrsfs}
\usepackage{subfigure}
\usepackage{multirow}
\usepackage{booktabs}
\usepackage{lscape}
\usepackage{thm-restate}
\usepackage{hyperref}
\usepackage{multirow}
\usepackage{float}
\usepackage{bm}
\usepackage{cleveref}
\usepackage{enumitem}

\usepackage[ruled,vlined]{algorithm2e}
\newtheorem{theorem}{Theorem}[section]

\newtheorem{proposition}[theorem]{Proposition}

\newtheorem{exam}{Example}

\DeclareMathOperator*{\argmin}{arg\,min}

\def\Sigmahat{\widehat{\Sigma}}

\newcommand{\hGam}{\mbox{$\widehat{\Gamma}$}}

\def\tr{\hbox{\rm tr}}
\def\supp{\hbox{\rm supp}}

\def\R{{\mathbb{R}}}
\def\calN{{\mathscr N}}
\def\calS{{\mathscr S}}\def\calA{{\cal A}}
\def\E{{\mathbb{E}}}

\newcommand{\Span}{\mathrm{Span}}
\def\var{\mbox{var}}

\def\vec{\mbox{vec}}

\def\wh{\widehat}
\def\cs{\mathcal{S}_{Y\mid X}}

\def\T{{ \mathrm{\scriptscriptstyle T} }}
\def\F{{ \mathrm{\scriptscriptstyle F} }}
\def\frechet{{Fr\'{e}chet }}

\newcommand\independent{\protect\mathpalette{\protect\independenT}{\perp}}
\def\independenT#1#2{\mathrel{\rlap{$#1#2$}\mkern2mu{#1#2}}}

\newcommand\norm[1]{\left\lVert#1\right\rVert}

\graphicspath{{figures/}}
\begin{document}

\title{\bf Sparse Fr\'{e}chet Sufficient Dimension Reduction with Graphical Structure Among Predictors}
  \author{Jiaying Weng \\
  	Department of Mathematical Sciences, Bentley University,\\
	Kai Tan\\
	Department of Statistics, Rutgers, The State University of New Jersey,\\
    Cheng Wang \\
    School of Mathematical Sciences, Shanghai Jiaotong University,\\
	and \\
    Zhou Yu\\
    School of Statistics, East China Normal University\\
	}
	\date{}
  \maketitle

\begin{abstract}
    Fr\'{e}chet regression has received considerable attention to model metric-space valued responses that are complex and non-Euclidean data, such as probability distributions and vectors on the unit sphere.
	However, existing \frechet regression literature focuses on the classical setting where the predictor dimension is fixed, and the sample size goes to infinity. 
	This paper proposes sparse \frechet sufficient dimension reduction with graphical structure among high-dimensional Euclidean predictors.
	In particular, we propose a convex optimization problem that leverages the graphical information among predictors and avoids inverting the high-dimensional covariance matrix. 
 We also provide the Alternating Direction Method of Multipliers (ADMM) algorithm to solve the optimization problem. 
 Theoretically, the proposed method achieves subspace estimation and variable selection consistency under suitable conditions.
  Extensive simulations and a real data analysis are carried out to illustrate the finite-sample performance of the proposed method. 
\end{abstract}
\begin{flushleft}
\end{flushleft}

\section{Introduction}\label{sec:intro}
\subsection{Background}

 Regression models with general types of responses have recently received extensive attention from statisticians and field experts due to their wide applications in real-world problems. See, for instance, various regression models for images, shapes, tensors, and densities \citep{peyre2009manifold,small2012statistical,li2017parsimonious,petersen2016functional}. 
Recently, \cite{petersen2019frechet} introduced Fr\'{e}chet regression model with random object response in a metric space and predictor in Euclidean space. 
\cite{tucker2021variable} further proposed a variable selection method for \frechet regression for metric-space valued responses. \cite{bhattacharjee2021single} investigated a single index Fr\'{e}chet regression model and interpreted the index as the direction in the predictor space along which the variability of the response is maximized. 

Sufficient dimension reduction (SDR) is a natural extension of single index models to multiple index models.
Let $X$ be a $p$-dimensional random vector in Euclidean space $\R^p$, and $Y$ be a random object in a metric space $(\Omega, m)$, where $m(\cdot, \cdot)$ is a metric. 
The primary goal of SDR is to reduce the dimensionality of a set of predictors $X$ while retaining as much information as possible about the response variable 
$Y$. 
To be specific, SDR aims to seek a $p \times d$ matrix $\bm\beta$, typically $d$ is much smaller than $p$, such that 
\begin{equation}\label{eq1}
    Y \independent X \mid \bm\beta^\T X.
\end{equation}
This means that the response $Y$ is statistically independent of $p$-dimensional predictor vector $X$ given the linear combination of predictors $\bm\beta^\T X$. 
In other words, all the information of $Y$ on $X$ is contained in the $d$-dimensional linear projection $\bm\beta^\T X$. 
In the classical linear regression model $Y = X \beta + \varepsilon$, the coefficient $\beta$ satisfies \eqref{eq1}, and the regression task is to estimate $\beta$ to recover the conditional mean $\E(Y \mid X) = X \beta$. The conditional independence \eqref{eq1} generalizes the linear model to more general regression models by allowing more flexible forms of $\E(Y \mid X)$ that are not restricted to linear form and can depend on multiple coefficient vectors. 
If we find $\bm\beta$ such that \eqref{eq1} holds, then 
the regression problem reduces to a lower-dimensional problem, because all of the regression information of $Y$ on $X$ can be summarized in the linear combinations $\bm\beta^\T X\in \R^d$.
This way, the dimension of predictor $p$ reduces to $d$, the dimension of $\bm\beta^\T X$, often referred to as the structural dimension in SDR literature. 
It's worth noting that the matrix $\bm\beta$ satisfying \eqref{eq1} is not identifiable since \eqref{eq1} still holds if replacing $\bm\beta$ with $\bm\beta A$ for any $ d \times d $ non-singular matrix $A$. On the other hand, 
the column space of $\bm\beta$, denoted by $\Span(\bm\beta)$, is identifiable. 
And the parameter of interest in SDR is the central subspace $\cs$, defined as the smallest column space of $\bm\beta$ satisfying \eqref{eq1}. 
The first SDR method, sliced inverse regression (SIR), was introduced by \cite{li1991}. 
Since then, a variety of SDR approaches have been proposed, including sliced average variance estimation (SAVE; \citealt{cook1991}), principal Hessian direction (pHd; \citealt{li1992principal}), cumulative mean estimation (CUME; \citealt{zhu2010dimension}), the Fourier transform approach \citep{weng2018fourier,weng2022yin}, and the measure-based approach \citep{sheng2016sufficient,sheng2023dimension}. For further concepts about SDR, we refer to the review articles \citep{review_sparsesdr,review_slicingfreesdr}  and related chapters in \cite{li2018sufficient}. 

While classical SDR approaches can not handle regression with responses in non-Euclidean spaces, 
\frechet SDR methods have recently been developed for metric-space valued responses and predictors in $\R^p$.
\cite{chao2022frechet} proposed a Fr\'{e}chet SDR approach with a novel kernel matrix by taking advantage of the metric distance  of the random objects. 
\cite{zhang2021dimension} considered transforming the \frechet regression into traditional regression by mapping the metric-space valued response to a real-valued response. 
These two methods are developed in the classical regime where the sample size $n$ is larger than the dimension of predictors $p$. 
They suffer from two issues for a high-dimensional regime where $p$ is much larger than $n$. First, their methods require computing the inverse of a $p\times p$ covariance matrix, which is a challenging problem when $p$ is greater than $n$ \citep{li2008sliced}. 
Second, the estimated linear combinations from their methods contain all the $p$-dimensional predictors, which often makes it difficult to interpret the extracted components \citep{tan2020sparse}.
Over the last two decades, many sparse SDR approaches have been proposed to achieve dimension reduction and variable selection simultaneously.
There are two lines of research:
one is to solve the original problem by solving a sequence of subproblems with lower dimensions \citep{yin2015sequential, yu2016marginal, yu2016trace}; and the other way is to construct an equivalent optimization problem without the inverse of covariance matrix \citep{li2008sliced, tan2018convex, lin2019sparse, qian2018sparse,tan2020sparse, tang2020high, weng2022fourier, zeng2022subspace}.
Despite the vast literature on sparse SDR, few consider graph or network data, which are commonly encountered in real problems, for instance, gene data in biological studies and fMRI data in clinical and research activities.
The linear model has been extended to incorporate the graphical structure among predictors, see, for instance, \cite{li2008network, zhu2013simultaneous}. \cite{yu2016sparse} incorporated the predictor's structure node-by-node under the linear regression framework, which was extended to linear discriminant analysis in \cite{liu2019graph}.
Beyond the linear model, it will be interesting to develop SDR methods that can leverage the graphical information to improve the classical SDR method.

In this paper, we propose a unified sparse Fr\'{e}chet SDR framework for regression with metric-space valued responses and predictors with graphical structure in the high-dimensional regime where dimension $p$ can be much larger than sample size $n$. 
The main idea is to formulate the matrix of interest as a quadratic optimization problem and then employ a graph-based penalization to simultaneously achieve dimension reduction and variable selection. 
Our proposed framework can be applied to most SDR methods, including SIR, SAVE, CUME, pHd, and more.
Theoretically, we establish the consistency of subspace estimation and variable selection under proper conditions. 
For the computation, we provide an efficient ADMM algorithm to solve the optimization problem. Extensive simulations and a real data analysis are carried out to illustrate the superior performance of proposed estimators
\footnote{The python code for the simulation and the data analysis is available at \url{https://github.com/SDR-VariableSelection/FrechetSDR/}.}.

\subsection{Notation}
For a matrix $A = (a_{ij}) \in\R^{m\times n}$,
its Frobenius norm is $\|A\|_{\F} = (\sum_{i,j} a^2_{ij})^{1/2}$,
its spectral (operator) norm $\| A\|$ is the largest singular value of $A$,
its max norm is $\| A\|_{\text{max}} = \max_{i,j}| a_{ij}|$, 
and its infinity norm is $\| A\|_{\infty} =\max_{1\le i \le m}\sum_{j=1}^n| a_{ij}|$.
For a finite set $I$, let $|I|$ and ${I^{\mathrm{c}}}$ denote its cardinality and complement, respectively. 
For two index sets $I$ and $J$, $A_{IJ}$ stands for the $|I|\times|J|$ submatrix formed by $a_{ij}$ with $(i,j) \in I \times J$.
For any positive integer $k$, $[k]$ denotes the set $\{1, 2, \cdots, k\}$, and let ${A}_{I} = A_{I [n]}$ be the submatrix formed by $a_{ij}$ with $(i,j) \in I \times [n]$.
For $\calS \subseteq \{(i,j), i \in [m], j\in [n]\}$, let $A_\calS = \{a_{ij}, (i,j) \in \calS\}$. 
Throughout, $c_0,C_0, c_1, C_1, \cdots,$ are universal constants independent of $n$ and $p$, and can vary from place to place.

\subsection{Organization of the paper}
The rest of the paper is organized as follows. 
Section \ref{method} describes a weighted inverse regression ensemble and introduces a convex optimization problem incorporating graphical structure among predictors. 
In Section \ref{algorithm}, we propose an ADMM algorithm to solve the convex optimization problem. 
Section \ref{theorem} establishes subspace estimation and variable selection consistency of the proposed method. 
In Section \ref{simulation}, numerical comparisons demonstrate the necessity of including the graphical structure and show the merit of the proposed loss function. The proofs of theoretical results are given in Appendix C.

\section{Fr\'{e}chet SDR with graphical structure}\label{method}

\subsection{Preliminary}
Consider Fr\'{e}chet regression with a response variable $Y$ in a metric space equipped with metric $m(\cdot,\cdot)$ and $p$-dimensional predictor vector $X$.
Let us mention a few quick examples of such responses:
(i) For responses in Euclidean space $\R^q$, the metric $m(y, \tilde{y}) = \norm{y-\tilde{y}}$ is the Euclidean distance, and the regression problem becomes multiple response regression. 
(ii) For responses that live in a unit sphere, then $m(y, \tilde{y}) = \arccos(y^\T \tilde{y})$ is the geodesic distance between $y$ and $\tilde{y}$.
(iii) For responses that are distribution functions, one metric for such metric space is the 2-Wasserstein distance, defined as $m(y, \tilde{y}) = [{\int_{0}^{1} \{y^{-1}(t) - \tilde{y}^{-1}(t)\}^2 dt}]^{1/2}$, where $y$ and $\tilde y$ are two distribution functions, and $y^{-1}(\cdot), \tilde y^{-1}(\cdot)$ are the corresponding quantile functions.

Classical SDR aims to find a $p \times d$ matrix $\bm\beta = [\beta_1,\cdots,\beta_d]$ such that $Y \independent X \mid \bm\beta^\T X$. 
Most SDR methods are often equivalent to the generalized eigenvalue problem $\Lambda \beta_i = \lambda_i \Sigma \beta_i, (i = 1, 2, \ldots, d)$ \citep{li2018sufficient}, where $\Lambda$ is a method-specific kernel matrix and $\Sigma = \var(X)$ is the covariance matrix of predictors. 
Furthermore, $\beta_1,\cdots,\beta_d$ can be cast as the first $d$ leading eigenvectors of $B=\Sigma^{-1} \Lambda \Sigma^{-1}$. 
Hence, the row and column sparsity of $B$ is equivalent to the row sparsity of $\beta_i, (i = 1, 2, \ldots, d)$.
When the linearity condition holds, namely, $\E(X \mid \bm\beta^\T X)$ is linear in $X$,
we have $\Span(\bm\beta) = \Span(\beta_1,\cdots,\beta_d)\subseteq\cs$ \citep{cook1998regression}.
In addition, if the coverage condition that $ \Span(\beta_1,\cdots,\beta_d) = \cs$ holds, then $(\beta_1, \cdots, \beta_d)$ constitutes a basis of $\cs$.

The concept of the kernel matrix, as explored in this paper, originates from \cite{shao2014martingale}, \cite{mai2023slicing}, and \cite{chao2022frechet}. In \cite{shao2014martingale}, the martingale difference divergence matrix (MDDM) was introduced to quantify the independence between two random vectors.
The definition of MDDM is as follows:  
\begin{align*}
    - \E[\{Y- \E(Y)\}^\top \{\widetilde{Y} - \E(\widetilde{Y})\}\|X - \widetilde{X}\|],
\end{align*}
where $(\widetilde{X}, \widetilde{Y})$ is an independent copy of $(X, Y)$, and both are random vectors.
While classical SDR methods only consider responses in Euclidean space, 
\cite{chao2022frechet} recently proposed
\frechet SDR to handle responses in a more general metric space. This literature introduced a weighted inverse regression ensemble (WIRE) with the following kernel matrix,
\begin{equation}\label{eq:Lambda}
	\Lambda = -\E\{ (X-\mu)(\widetilde{X} - \mu)^\T m(Y, \widetilde{Y})\} \in \R^{p\times p},
\end{equation}
where $Y, \widetilde{Y}$ are independent random objects and 
$\mu = \E(X)$. 
Compared to MDDM, the WIRE matrix switches the roles of response and predictor vectors to handle random objects in metric space.
In the above definition, the weight $m(Y, \widetilde Y)$ is the distance metric between $Y$ and $\widetilde Y$. 
This metric makes it possible to conduct SDR for response in general metric space.
We in this paper extend WIRE to estimate $\cs$ in the high-dimensional setting. To this end, we first need to estimate $B = \Sigma^{-1}\Lambda \Sigma^{-1}$. 
Once we obtain a norm consistent estimate $\widehat{B}$ of $B$, we can extract its $d$ leading eigenvectors $ (\hat{\beta}_1,\cdots,\hat{\beta}_d)$. Let $\wh{\bm\beta} =[\hat{\beta}_1,\cdots,\hat{\beta}_d]$, then the estimated central subspace is the column space of $\wh{\bm\beta}$. In the next subsection, we extend the WIRE estimator to handle high-dimensional predictors with graphical structure and provide a unified framework to obtain such $\widehat{B}$ by solving a convex optimization.

\subsection{Graphical weighted inverse regression ensemble}
When the dimension of predictors $p$ is large, we assume that only few predictors are relevant to the response. Sufficient variable selection aims to identify a subset $S \subseteq [p]$, such that $Y \independent X \mid X_S$. 
Our goal is to simultaneously achieve dimension reduction and variable selection by finding a row-sparse matrix $\bm\beta$. As mentioned before, we need first estimate $B = \Sigma^{-1} \Lambda \Sigma^{-1}$. 
A natural idea is to plug in the sample estimates of $\Sigma^{-1}$ and $\Lambda$, however, estimating $\Sigma^{-1}$ is a challenging problem in high-dimensional setting \citep{li2008sliced}. To circumvent this problem, 
we provide a penalized optimization problem to estimate $B$ efficiently. 
Let $\vec(B)$ denote the vectorization of the matrix $B$ and $\otimes$ denote the Kronecker product. We consider the quadratic function,
\begin{align}\label{eq:penalty}
	f(B) = \vec(B)^\T (\Sigma \otimes \Sigma) \vec(B)/2-\vec(B)^\T \vec(\Lambda).
\end{align}
The gradient of $f(B)$ is $\nabla f(B) = (\Sigma\otimes \Sigma)\vec(B) - \vec(\Lambda)$. Setting it to zero yields the result $\vec(B) = (\Sigma\otimes \Sigma)^{-1}\vec(\Lambda)$. This motivates us to consider the optimization problem: $\arg\min_{B} f(B)$.
The advantage of minimizing $f(B)$ is that it does not require the inversion of $\Sigma \otimes \Sigma$.

Given data $\{(x_i, y_i),i=1,\ldots,n\}$, we first estimate $\Sigma$ and $\Lambda$ by their sample counterparts 
$\widehat \Sigma=n^{-1}\sum_{i=1}^n \left(x_i - \bar{x}\right) \left(x_i - \bar{x}\right)^\T$, 
and $\widehat{\Lambda} = -\{n(n-1)\}^{-1} \sum_{1\leq i \neq j \leq n} (x_i - \bar{x})(x_j - \bar{x})^\T m(y_i, y_j)$,
where $\bar{x}=\frac{1}{n}\sum_{i=1}^n x_i$. 
To obtain a sparse estimate, we rewrite \eqref{eq:penalty} into matrix form, replace $\Sigma$ and $\Lambda$ by their sample estimates, and impose a sparse-inducing penalty on $B$. Therefore, we propose a penalized optimization problem,
	\begin{align}\label{pdr}
		\widehat B=\argmin_{\Theta  \in \R^{p \times p}} \frac{1}{2}\tr(\Theta^\T\Sigmahat \Theta\wh \Sigma  - 2 \Theta^\T\wh \Lambda) +\lambda g(\Theta),
	\end{align}
where $g(\cdot)$ is a convex penalty function whose strength is tuned by the parameter $\lambda>0$.  
This optimization problem avoids directly estimating $\Sigma^{-1}$ and can be efficiently solved by the ADMM algorithm presented in Section \ref{algorithm}. 
    The penalty function can be problem-specific, for example, \cite{tang2020high} utilized $\ell_1$ norm penalty to detect high-dimensional interactions; \cite{zeng2022subspace} imposed double penalties to detect structure dimension and select variables automatically. For high-dimensional predictors with graphical structures, we propose a novel penalty function that can leverage the graphical information and deal with correlated predictors.

\subsection{Implement the graphical penalty}
For high-dimensional linear regression, 
\cite{yu2016sparse} considered a graph-based penalized method to incorporate graphical structure among predictors. 
In this section, we leverage the graphical information in a model-free setting using SDR. 
To our knowledge, this is the first attempt to incorporate graphical structure in the SDR framework. 
        
When $X = (X_1, \ldots, X_p)^\T$ is a Gaussian vector with covariance matrix $\Sigma$. A well-known result in graphical models states that 
$X_i$ and $X_j$ are not connected if and only if $X_i$ and $X_j$ are independent given $\{X_k, k\in[p], k\ne i,j\}$ \citep{lauritzen1996graphical}.
For Gaussian vector $X$, 
the conditional independence between $X_i$ and $X_j$ given other variables is equivalent to $\omega_{ij}=0$, where $\omega_{ij}$ is the $(i,j)$ entry of $\Omega = \Sigma^{-1}$. 
To utilize the graphical information among predictors $X$, 
we first write $B = \Omega \Lambda \Omega=\Omega A$ with $A = \Lambda \Omega$. 
Let $\omega_{i} = (\omega_{i1}, \cdots, \omega_{ip})^\T$ be the $i$th row of $\Omega$, $a_i\in \R^p$ be the $i$th row of $A$, and $Q^{(i)} = \omega_{i} a_i^\T\in \R^{p\times p}$.
Thus, $B = \sum_{i=1}^{p} \omega_{i} a_i^\T = \sum_{i = 1}^{p} Q^{(i)}$.
If $X_i$ is not connected with $X_j$, then $\omega_{ij}=\omega_{ji}=0$, hence the $j$th row of $Q^{(i)}$ will be all zeros. In other words, each $Q^{(i)}$ is a row-sparse matrix. 
However, the decomposition of the matrix $B$ is not unique.
As $B$ is symmetric, its row sparsity is equivalent to its column sparsity. Therefore, it is always possible to find a decomposition where each component is both row-sparse and column-sparse.

Without loss of generality, we assume that there is a latent decomposition of $B$ into $p$ matrices $V^{(1)}, V^{(2)}, \cdots, V^{(p)} \in \R^{p\times p}$ such that $B = \sum_{i = 1}^{p} V^{(i)}$, where $V^{(i)}$ are row- and column-sparse. 
Set $S$ is defined as the index set of active variables, i.e., 
$S = \{i\in[p]: B_{i} \neq 0 \}$, which is the index of non-zero rows and columns of $B$.
Instead of estimating $B$ directly, we propose an optimization problem that minimizes the objective function over $V^{(i)}$, for all $i \in [p]$ and impose penalties on each $V^{(i)}$ to induce a sparse estimate. 
Let set $N_i$ be the index set of the neighborhood of the $i$th predictor, i.e., $N_i = \{k: \omega_{ki}\neq 0, 1\leq k \leq p\}$.
The desired penalty term should shrink some $V^{(i)}$ to $0$, and force the nonzero entries of other $V^{(i)}$s belong to $ \calN_{i}= N_i \times S, (i = 1, 2, \cdots, p)$. 
Hence, we consider each matrix $V^{(i)}$ as a group and propose a graphical penalty 
\begin{align*}
	\|\Theta\|_{G, \tau} = \min_{\sum_{i=1}^{p}V^{(i)} = \Theta,\  \supp(V^{(i)}) \subseteq \calN_i} \sum_{i=1}^p \tau_i\|V^{(i)}\|_{\F},
\end{align*}
where $\supp(U) = \{(i,j): U_{ij}\neq 0\}$ and $\tau_i>0$. 
This function $\|\Theta\|_{G, \tau}$ is a norm, therefore convex \citep{obozinski2011group}, 
so many algorithms can be used to solve the optimization problem with this new penalty. 

To utilize graphical information,
we consider the optimization problem (4) but taking the penalty term $g(\Theta)$ as the graphical penalty $\|\Theta\|_{G,\tau}$:
\begin{align}\label{obj}
	\widehat{B} = &\argmin_{\Theta\in\R^{p\times p}}\frac{1}{2}\tr\{\Theta^\T (\Sigmahat \Theta\wh \Sigma  - 2 \wh \Lambda)\} + \lambda \|\Theta\|_{G, \tau},
\end{align}
In this context, $\widehat{B}$ represents the  minimizer of the population value $B$, with $\Theta$ serving as a variable in the objective function. 
This optimization problem, \eqref{obj}, can be reformulated as follows and solved using the ADMM algorithm:
\begin{align}\label{obj2}
    &\min_{\Theta, V^{(1)}, \cdots, V^{(p)}}\frac{1}{2}\tr\{\Theta^\T (\Sigmahat \Theta \Sigmahat  - 2 \wh\Lambda)\} + \lambda \sum_{i=1}^p \tau_i\|V^{(i)}\|_{\F},\\
    &\qquad\text{s.t. } \quad\sum_{i=1}^{p}V^{(i)} = \Theta,\quad \text{and}\quad \supp(V^{(i)}) \subseteq \calN_i, i \in [p].\nonumber
\end{align}
The constraints include $\Theta = \sum_{i = 1}^{p} V^{(i)}$ and the support of each $V^{(i)}$ is in $\calN_i \subseteq S$. Notably, $S$ plays a role in the optimization problem through its influence on $\calN_i$.
Proposition \ref{prop:02} in Section \ref{theorem} provides a sufficient and necessary condition for the existence of the solution. This condition shows us a way to construct a solution to the optimization problem. The decomposition components $\{V^{(i)}\}$ are not unique, but the estimate $\widehat{B}$ should be unique.
We refer to the minimizer of \eqref{obj2} as \textbf{G}raphical  \textbf{W}eighted \textbf{I}nverse \textbf{R}egression \textbf{E}nsemble (GWIRE).

Our proposed method is a distinct SDR method capable of addressing non-Euclidean or multivariate responses. 
\cite{chao2022frechet} proposed a WIRE estimator for Fr\'echet SDR in low-dimensional cases. 
\cite{mai2023slicing} utilized MDDM to deal with multivariate responses in high-dimensional scenarios. 
Our methodology represents a significant extension of these foundational works, particularly in its application to high-dimensional predictors that exhibit graphical structures within the framework of Fr\'echet SDR. This extension is not only innovative but also critical in accommodating the complexities associated with high-dimensional variables.
Furthermore, the proposed optimization problem can be directly extended to other SDR methods by substituting the relevant kernel matrices $\Lambda$ into the optimization problem \eqref{obj}.
For example, if the WIRE kernel matrix $\Lambda$ in \Cref{obj} is replaced by other SDR kernel matrices, such at SIR, SAVE, CUME, and pHd, then it is able to solve the corresponding SDR problems. Hence, the proposal is a general framework for SDR with graphical structure.
In the real data scenarios, the graphical structure of predictors is usually unknown.
In this case, we can estimate the precision matrix $\Omega$ by the graphical lasso \citep{friedman2008sparse} and then obtain the neighborhood information.

\section{Estimation procedure}\label{algorithm}


\subsection{ADMM Algorithm for GWIRE}
This section employs the ADMM algorithm \citep[Section 6.3]{boyd2011distributed} to solve the optimization problems \eqref{obj2}.
ADMM algorithm has found its applications in recent SDR literature, see, for example, \cite{tan2020sparse,tang2020high,weng2022fourier, zeng2022subspace}.
The principle of ADMM is to convert the original problem into a sequence of sub-problems involving only the smooth loss function or the convex penalty function. 

We first rewrite optimization problem \eqref{obj2} into an augmented Lagrangian dual problem. 
\begin{align}\label{algo:gwire}
  \mathcal{L}_{\rho}(\Theta, \{V^{(i)}\}^p_{i=1}, W)  = \frac{1}{2}\tr\{\Theta^\T (\Sigmahat \Theta\wh \Sigma  - 2 \wh \Lambda)\}  + \lambda \sum_{i=1}^{p} \tau_i\|V^{(i)}\|_{\F} + \frac{\rho}{2} \|\Theta+W - V\|_{\F}^2,
\end{align}
where $V = \sum_{i=1}^p V^{(i)}$, $W \in \R^{p\times p}$ is the dual variable, and $\rho$ is a positive penalty parameter.
Based on the ADMM algorithm, it is necessary to introduce different notations to distinguish $\Theta$ from $\sum_{i=1}^p V^{(i)}$. Consequently, we use $V$ to represent the summation of $V^{(i)}$. This notation aids in updating the variables $\Theta$ and $V^{(i)}$ separately.
Given initial values for each $V^{(i)}$ and $W$, we solve \eqref{algo:gwire} by iterating the following three steps:
\begin{align}
	\Theta^{k+1} & = \argmin_{\Theta: \Theta = \Theta^{\T}} \mathcal{L}_{\rho}(\Theta, \{V^{(i),k}\}^p_{i=1}, W^k) \label{eq:stepsB} \\
        V^{(i),k+1} & = \argmin_{V^{(i)}} \mathcal{L}_{\rho}(\Theta^{k+1}, \{V^{(j),k+1}\}_{j<i}, V^{(i)},\{V^{(j),k}\}_{j>i}, W^k), \text{ for each } i \in [p] \label{eq:stepsV}\\
         & \qquad \text{ s.t. }\quad  \{j \in [p]: \|V^{(i)}_j\|\neq 0\} \subseteq N_i \nonumber\\
	W^{k+1} & = W^k + \Theta^{k+1}- V^{k+1}, \text{ where } V^{k+1} = \sum_{i=1}^p V^{(i),k+1}. \label{eq:stepsW}
\end{align}
The above three steps are easy to implement.
The optimization problem \eqref{eq:stepsB} is strongly convex since the Hessian of $\mathcal{L}_{\rho}(\Theta, V,W)$ with respect to $\Theta$ is positive definite, i.e., $\frac{\partial^2 \mathcal{L}_{\rho}}{\partial \Theta\partial \Theta^\T} = \wh \Sigma  \otimes \wh \Sigma + \rho  I \succ 0.$
Thus, the problem \eqref{eq:stepsB} has a unique solution, which can be derived using Proposition \ref{thm:admm}.
\begin{proposition}\label{thm:admm}
    Consider the following optimization problem: 
    \begin{align*}
        \wh \Theta = \argmin_{\Theta\in \R^{p\times p}} \frac{1}{2}\tr\{\Theta^\T (\Sigma^* \Theta \Sigma^*  - 2 \Lambda^*)\} + \frac{\rho}{2}\|\Theta +W^* - V^*\|^2_{\F},
    \end{align*}
    where $V^*, W^*$, and $\Lambda^*$ are symmetric matrices of dimension $p$, and $\Sigma^*$ is a positive semi-definite matrix of the same dimension. 
    Let $\Sigma^* = P\Gamma P^\T$ be the spectral decomposition of $\Sigma^*$, where $\Gamma$ is a diagonal matrix with ordered entries $\gamma_1\ge\gamma_2\ge\ldots\ge\gamma_p$ on its diagonal. 
    Then the closed-form solution for $\wh\Theta$ is given by:
    \begin{align}\label{eq:B-closedform}
        \wh \Theta = L - P\{C\circ (P^\T L P)\}P^\T,
    \end{align}
    where $L = \Lambda^*/\rho + V^* - W^*$, $C$ is a $p\times p$ matrix with entries $C_{ij} = {\gamma_i \gamma_j}/(\gamma_i\gamma_j + \rho)$, and $\circ$ denotes the Hadamard product.
\end{proposition}
When solving \eqref{eq:stepsB} in the $(k+1)$th iteration, we use Proposition \ref{thm:admm} to solve for $\Theta^{k+1}$ by replacing $\Sigma^*$ with the sample covariance matrix $\wh\Sigma$, $\Lambda$ with sample $\wh\Lambda$, $V^*$ with the symmetrized matrix $\{V^{k} + (V^{k})^{\T}\}/2$, and $W^*$ with $W^k$ from previous step.
To solve \eqref{eq:stepsV}, by first-order optimality conditions, we have the following closed-form expression of 
$V^{(i),k+1}$,
\begin{equation}\label{eq:admmV}
	\begin{array}{lll}
		V^{(i),k+1}_{N_i} & = & 
   \bigl(1 -\frac{\lambda \tau_i}{\rho \|U^{i,k+1}_{N_i} \|_{\F}}\bigr)_{+} U^{i,k+1}_{N_i},\\
   V^{(i),k+1}_{N_i^c} & = & 0,
	\end{array}
\end{equation}
where $(a)_{+} = \max\{a, 0\}$ and $U^{i,k+1} = \Theta^{k+1} + W^{k} - \sum_{j<i} V^{(j),k+1} - \sum_{j>i} V^{(j),k}$. 
Updating $W$ in \eqref{eq:stepsW} is straightforward because it is just a simple operation of matrices.
In summary, the ADMM algorithm for solving \eqref{obj2} is presented below.

\begin{algorithm}[H]
	\label{algo1}
	\SetAlgoLined
    \textbf{Input:} $\lambda$, $\rho$, $\widehat{\Sigma}$, and $\widehat{\Lambda}$. \\
	\textbf{Initialization:} $W^0$, $V^{(i),0}$, $V^0 = \sum_{i=1}^p V^{(i),0}$
    and $k=0$. \\
    Let $G = \{i: |N_i| > 1\}$ and $G^c = [p]\setminus G$.\\
	\Repeat{$\|\Theta^{k} - \Theta^{k-1}\|_{\F} \leq \epsilon^{\mathrm{pri}} $ and $\rho\|V^{k} - V^{k-1}\|_{\F} \leq \epsilon^{\mathrm{dual}}$}{
    \underline{update $\Theta$:}\\
    $\Theta^{k+1} \gets L - P\{C\circ (P^\T L P)\}P^\T$, where $L = \wh \Lambda/\rho + \{V^{k} + (V^{k})^{\T}\}/2 - W^k$, and $C$, $P$ are defined in Proposition \ref{thm:admm}. \\
    \underline{update $V$:}\\
    \For{$i \in G$}{
        $U^{i,k+1} \gets \Theta^{k+1} + W^{k} - \sum_{j<i} V^{(j),k+1} - \sum_{j>i} V^{(j),k}$,\\
        $V^{(i),k+1}_{N_i} \gets \left(1 -\frac{\lambda \tau_i}{\rho \|U^{i,k+1}_{N_i} \|_{\F}}\right)_{+} U^{i,k+1}_{N_i}$. 
    }
    $U^{k+1}_{G^c} \gets \Theta^{k+1}_{G^c} + W^{k}_{G^c}$.\\
    \For{$j \in G^c$}{
        $V^{(j),k+1}_j \gets \left(1 -\frac{\lambda \tau_j}{\rho \|U^{k+1}_{j} \|_{2}}\right)_{+} U^{k+1}_{j}$.
    }
    \underline{update $W$:}\\
    $W^{k+1} \gets W^{k} + \Theta^{k+1} - \{V^{k+1} + (V^{k+1})^{\T}\}/2$.\\
    $k \gets k + 1$.\\
	}
    
	Let $\widetilde{B} = V^{k}$, and  $\widehat{S} = \{i: \widetilde{B}_i \neq 0\}$, then set the columns of $\widetilde{B}$ in $\widehat{S}^c$ to 0. \\
	\textbf{Output:} $\wh B = (\widetilde{B} + \widetilde{B}^{\T})/2$ and $\widehat{S}$. 
	\caption{Algorithm for solving GWIRE.}
\end{algorithm}

Algorithm~\ref{algo1} is easy to implement because each subproblem has a closed-form solution.
The ADMM algorithm iterates between $\Theta$, $V$, and $W$. To solve $\Theta$, we propose an explicit expression in Proposition \ref{thm:admm}, involving only matrix operations. For the solution of $V$, we divide the problem into two steps: one for $N_i>1$ when the $i$-th predictor has more than one neighbor, and the other for $N_i=1$ when the $i$-th predictor only  have one neighbor, that is itself. The second step can be solved without the need for a loop.
For each iteration, the computational complexity is of order $p^3$ due to the multiplication of three $p\times p$ matrices in $P^\T L P$.
The stopping criteria are that the differences between the current and previous iterations are smaller than thresholds. We adopt the criteria from \cite{boyd2011distributed}. The iteration is stopped when
$\|\Theta^{k} - \Theta^{k-1}\|{\F} \leq \epsilon^{\mathrm{pri}}$ and $ \rho \|V^{k} - V^{k-1}\|\F \leq \epsilon^{\mathrm{dual}},$
where $\epsilon^{\mathrm{pri}}$ and $\epsilon^{\mathrm{dual}}$ are feasibility tolerances for the primal and dual feasibility conditions. We set $\epsilon^{\mathrm{pri}} = \epsilon^{\mathrm{dual}} = 10^{-3}$ in our simulation studies.

\subsection{Implementation details}

When implementing Algorithm \ref{algo1}, 
we take $\rho=1$, $V^{(i)}=0$, and $W=0$. 
For the weight $\tau_i$ in \eqref{obj2}, we use $\tau_i = \sqrt{|N_i|}$ to take into account the  number of neighbors for each variable.
We now provide a data-driven method to choose the tuning parameter $\lambda$ and the structural dimension $d$.  

To select structural dimension, we employ the ladle estimator for order determination proposed by \cite{luo2016combining}. This approach determine the rank of a matrix by combining the patterns of eigenvalues and eigenvectors. 
The ladle estimator is obtained by constructing a ladle plot, which is a combination of the scree plot of sample eigenvalues and the bootstrap variability plot of eigenvectors. 
Theorem 2 in \cite{luo2016combining} provided a theoretical guarantee of the reliability and consistency of the ladle estimator in determining the rank of a matrix. This theoretical result supports the practical effectiveness of the ladle estimator and enhances its credibility as a robust method for order determination in statistical analysis.
We first fix $\lambda = \lambda_{\max}/5$ to estimate $\wh B$ by Algorithm \ref{algo1}, where $\lambda_{\max}= \max\{\|\wh\Lambda_{N_i}\|_{\F}/\tau_i, i = 1,\cdots, p\}$ is the maximum $\lambda$ that leads to an all-zeros $\wh B$. 
Then, generate 100 bootstrap sample and obtain  
the bootstrap estimates $\{\wh B^b: b = 1,..., 100\}$. 
Define the eigenvalues and eigenvectors for $\wh B$ as $\{\hat\phi_1, \cdots, \hat\phi_p, \hat\beta_1, \cdots, \hat\beta_p\}$, and similarly for $\wh B^{b}$ as $\{\hat\phi_1^{b}, \cdots, \hat\phi_p^{b}, \hat\beta_1^{b}, \cdots, \hat\beta_p^{b}\}$.
For each $k < |\wh S|$, let $\hat{\bm\beta}(k) = [\hat\beta_1, \cdots, \hat\beta_k]$ and $\hat{\bm\beta}^{b}(k) = [\hat\beta_1^{b}, \cdots, \hat\beta_k^{b}]$. 
We further define two functions from $\{0, \cdots, |\wh S|-1\}$ to $\R$, 
$f(k) = f^0(k) / \{1 + \sum_{i = 0}^{|\wh S| -1} f^0(i)\}$ and $h(k) = \hat\phi_{k+1}/(1 + \sum_{i = 0}^{|\wh S|-1} \hat\phi_{i+1})$, where
\[
    f^0(k)= 
\begin{cases}
    0,& \text{if } k=0,\\
    100^{-1}\sum^{100}_{b = 0} [1 - |\det\{\hat{\bm\beta}^{\T}(k)\hat{\bm\beta}^{b}(k)\}|],              & \text{if } k = 1, \cdots, |\wh S|-1.
\end{cases}
\] 
At last, the estimate of $d$ minimizes $f(k) + h(k)$, 
\begin{equation}\label{eq:d-hat}
    \hat{d} = \argmin \{f(k) + h(k): k = 0, \cdots, |\wh S|-1\}.
\end{equation}

We select the tuning parameter $\lambda$ by $10$-fold cross-validation using $d$ (or $\hat d$ if $d$ is unknown). 
For the candidate set of $\lambda$, we consider a
sequence of $30$ log-spaced values from $0.05\lambda_{\max}$ to $\lambda_{\max}$.
For each $k = 1, \cdots, 10$, solve the optimization problem \eqref{obj2} for each $\lambda$ and denote the solution by $\wh B^{-k}(\lambda)$, and the sample covariance matrix by $\wh\Sigma^{-k}$ using all the observations outside the $k$th fold. 
Then obtain the estimated directions by spectral decomposition of $\wh B^{-k}(\lambda)$. Let $\hat \beta_1^{-k}(\lambda), \ldots, \hat\beta_p^{-k}(\lambda)$ denote the sorted eigenvectors of $\wh B^{-k}(\lambda)$ corresponding to decreasing eigenvalues, we further use the structural dimension $d$ to obtain the useful directions as $\hat{\bm\beta}^{-k}(d;\lambda)=[\hat \beta_1^{-k}(\lambda), \ldots, \hat\beta_{ d}^{-k}(\lambda)]$.
We also standardize $\hat{\bm\beta}^{-k}( d;\lambda)$ such that $\{\hat{\bm\beta}^{-k}(d;\lambda)\}^{\T} \wh\Sigma^{-k} \hat{\bm\beta}^{-k}(d;\lambda) = I_{d}$ by setting
$
    \hat{\bm\beta}^{-k}( d;\lambda) = \hat{\bm\beta}^{-k}( d;\lambda)\left[\hat{\bm\beta}^{-k}( d;\lambda)^\T\hat\Sigma^{-k} \hat{\bm\beta}^{-k}( d;\lambda)\right]^{-1/2}.
$
Finally, we choose the $\lambda$ that minimize 
the following criterion from \cite{chen2010coordinate}, 
\begin{align*} \frac{1}{10}\sum^{10}_{k=1}-\tr[\{\hat{\bm\beta}^{-k}(d;\lambda)\}^\T \wh \Lambda^{k} \hat{\bm\beta}^{-k}(d;\lambda)].
\end{align*}

\section{Theoretical results}\label{theorem}

In this section, we show subspace estimation and variable selection consistency for GWIRE. 
To be specific, Theorem \ref{thm1} states non-asymptotic estimation error bound of $\wh B$ and variable selection consistency. 
Theorem \ref{thm2} further provides the rate of convergence between estimated and true central subspace under the Frobenius norm.
To begin with, we provide the optimality conditions for GWIRE in the following proposition.
\begin{proposition}\label{prop:02}
	A symmetric matrix $\wh B \in \R^{p \times p}$ is a solution of \eqref{obj2} if and only if $\wh B$ can be decomposed as $\wh B = \sum_{i=1}^p V^{(i)}$, where each $V^{(i)}$ satisfies 
  (a) $V^{(i)}_{\calN_i^c} = 0$; (b) either $V^{(i)}_{\calN_{i}} \neq 0$ and $(\widehat{\Lambda} - \widehat{\Sigma}\wh{B}\widehat{\Sigma})_{\calN_{i}} = \lambda {\tau_i V^{(i)}_{\calN_i}}/{\|V^{(i)}_{\calN_i}\|_{\F}}$, or $V^{(i)}_{\calN_{i}} = 0$ and $\|(\widehat{\Lambda} - \widehat{\Sigma}\wh{B}\widehat{\Sigma})_{\calN_{i}} \|_{\F} \leq \lambda \tau_i$. 
\end{proposition}
Proposition~\ref{prop:02} is a direct consequence of the sub-gradient conditions for the latent group Lasso problem in \cite{obozinski2011group}[Lemma 11] and \cite{yu2016sparse}[Proposition 1]. 
This proposition outlines the sufficient and necessary conditions for resolving optimization problem \eqref{obj2} in both population and sample contexts. These conditions relate to the graphical structure of $B$. Specifically, condition (a) states that the nonzero rows in each $V^{(i)}$ match the neighbors of the $i$th predictor, while condition (b) confirms that the solution complies with the KKT conditions.
With the sub-gradient conditions, we are ready to investigate the error bound for GWIRE.

Recall $B = \Sigma^{-1}\Lambda \Sigma^{-1}$, denote by $\calS = \{(i,j): B_{ij} \neq 0\}$ the set of nonzero positions of $B$ and its cardinality by $s = |\calS|$.
Let $s^* = |S|$ be the number of nonzero rows of $B$, $\Gamma = \Sigma \otimes \Sigma$ with  $\Gamma_{(i,j), (k,l)} = \Sigma_{ik} \Sigma_{jl}$, and $\Gamma_{\calA_1, \calA_2} = \{\Gamma_{(i,j), (k,l)}, (i,j) \in \calA_1, (k,l) \in \calA_2\}.$
The following assumptions will be useful.

\begin{enumerate}[label = (A\arabic*)]
\item\label{A1:gaussian} The covariate vector $X=(X_1,\dots,X_p)^\T$ is a sub-Gaussian random vector, i.e., \\ $E\{ \exp (c_0 |e^\T X |^2) \}\le c_1 < \infty$ for any unit vector $e$.
\item \label{A2:neighbor} The neighborhood set $N_i \subseteq S$ for $i \in S$. 
\item\label{A3:irrepresent} The irrepresentability condition holds, i.e., $\|\Gamma_{\calS^c,\calS}\Gamma^{-1}_{\calS,\calS}\|_{\infty} < \alpha/\tau_{*}$,\\ where $\alpha = \min_{i \in S^c}{\tau_i}/{\sqrt{d_i s^*}}$ with $d_i = |N_i|$, and $\tau_{*} = \max_{i \in S} \tau_i$. 
\item\label{A4:regime} The asymptotic regime is $s (\log p/n)^{1/2} \|\Gamma_{\calS,\calS}^{-1}\|_{\infty} \rightarrow 0$ and $p \rightarrow \infty$ as $n \rightarrow \infty$.
\item\label{A5:kernel} The sample kernel matrix $\wh \Lambda$ is consistent in the sense that $pr\big( \|\wh \Lambda-\Lambda\|_{\text{max}} > f(n,p) \big)\to 0$, 
where $f(n,p)$ depends on $(n,~p)$ and goes to zero as $n \to \infty$. 
\end{enumerate}
Assumption \ref{A1:gaussian} is a common assumption in high dimensional regressions.
Assumption \ref{A2:neighbor} mandates that if an active predictor $X_i$ is connected to another predictor $X_j$ ($j \in N_i$), then $X_j$ must also be active. This is a necessary condition for the graphical structure and was similarly stipulated by \cite{yu2016sparse} for sparse regression involving graphical structure.
The irrepresentability condition \ref{A3:irrepresent} is pivotal for establishing the oracle property of our method, mirroring a similar condition in LASSO and group LASSO
\citep{buhlmann2011statistics}.
The asymptotic regime in Assumption \ref{A4:regime} requires $\log p /n \to 0$ if both sparsity $s$ and $\|\Gamma^{-1}_{\calS \calS}\|_{\infty}$ are bounded. 
This allows $p$ to grow exponentially with $n$. 
On the other hand, if the sparsity $s = c' p$ for some constant $c' \in (0,1)$ and $\|\Gamma^{-1}_{\calS \calS}\|_{\infty}\le C$ for some constant $C$ (this holds when the spectrum of $\Sigma$ is bounded). 
Then, Assumption \ref{A4:regime} reads  
$c' n^{\alpha} \sqrt{\frac{\log n^\alpha}{n}} = o(1)$, 
which is equivalent to $\frac{\log n}{n^{1-2\alpha}} = o(1)$, which holds when $\alpha < 1/2$. 
Therefore, if $p = O(n^\alpha)$ for $\alpha < 1/2$, then Assumption \ref{A4:regime} holds.
Assumption~\ref{A5:kernel} guarantees the sample kernel matrix is a good estimate of the population kernel matrix.

Let $~ \Delta = \|(\Gamma-\hGam)\vec{(B)}\|_{\text{max}}+\|\wh \Lambda-\Lambda\|_{\text{max}}$,
and $~\delta= s\|\Gamma_{\calS,\calS}^{-1}\|_{\infty}  \|\Gamma-\hGam\|_{\text{max}}$.
Theorem \ref{thm1} provides the error bound for $\wh B$ under the max norm and the variable selection consistency. 
\begin{theorem}\label{thm1}
	Assume Assumptions \ref{A2:neighbor} and \ref{A3:irrepresent} hold. For GWIRE estimator $\wh B$ in \eqref{obj2} with $
	\lambda >  \Delta  ({1+\|\Gamma_{\calS^c,\calS}\Gamma^{-1}_{\calS,\calS}\|_{\infty} })/\{\alpha- \tau_{*}\|\Gamma_{\calS^c,\calS}\Gamma^{-1}_{\calS,\calS}\|_{\infty} - (\alpha +\tau_{*})\delta\}
	$, if
	$\delta< \min\{({\alpha - \tau_{*} \|\Gamma_{\calS^c,\calS}\Gamma^{-1}_{\calS,\calS}\|_{\infty} })/({\alpha + \tau_{*}}), 1\}$,
	 then
	\begin{align}
		\wh B_{\calS^c}={0}, \quad \text{~and~}	\|\wh B-B\|_{\text{max}}  \leq \|\Gamma_{\calS,\calS}^{-1}\|_{\infty}  \frac{\tau_{*} + \alpha}{1 + \|\Gamma_{\calS^c,\calS}\Gamma^{-1}_{\calS,\calS}\|_{\infty} }\lambda.
	\end{align}
	If in addition the minimum value of the $B_{\calS}$ is bounded from below as $\min_{(i,j)\in \calS} |B_{ij}| > 2 \lambda \tau_{*} \|\Gamma_{\calS,\calS}^{-1}\|_{\infty}$, then the variable selection is consistent $\wh \calS = \calS$. 
\end{theorem}

Recall that the estimated directions $\wh{\bm{\beta}}= [\hat{\beta}_1,\ldots,\hat{\beta}_d]$ consists of the first $d$ leading eigenvectors of $\widehat{B}$. 
We adopt the general loss $\|\wh{\bm \beta}\wh{\bm \beta} ^\T - \bm \beta \bm \beta^\T\|_{\F}$ from \cite{tan2020sparse} to evaluate the distance between the central subspace and its estimate.
Theorem~\ref{thm2} guarantees the central subspace estimation and variable selection consistency. 
\begin{theorem}\label{thm2}
Assume Assumptions \ref{A1:gaussian}--\ref{A5:kernel} and $\min_{(i,j)\in \calS} |B_{ij}| > 2 \lambda \tau_{*} \|\Gamma_{\calS,\calS}^{-1}\|_{\infty}$ hold. For GWIRE estimator $\wh B$ in \eqref{obj2} with $\lambda = c_2\{(\log p/n)^{1/2}+f(n,p)\}$ for sufficiently large $c_2$, we have
	\begin{align*}
		 \quad \|\wh{\bm \beta}\wh{\bm \beta} ^\T - \bm \beta \bm \beta^\T\|_{\F}
        =   O_P[\phi_d^{-1} s_0\|\Gamma_{\calS,\calS}^{-1}\|_{\infty}  \{(\log p/n)^{1/2}+f(n,p)\}],\quad
          \mbox{~and~} \quad\widehat{S} = S,
	\end{align*}
 where $s_0$ is the maximum number of non-zero elements for each row of $B$, and $\phi_d$ is the $d$th largest eigenvalue of $B$. 
\end{theorem}
Theorems~\ref{thm1} and \ref{thm2} are stated for the GWIRE estimator using sample kernel matrix $\wh\Lambda$ for WIRE. If the response variable is a scalar in $\R$, classical SDR kernel matrices (for example, SIR, CUME, pHd, etc.) can be used to replace $\wh\Lambda$, Theorem~\ref{thm2} still holds provided that Assumption \ref{A5:kernel} can be verified. 
Under the asymptotic regime described in Assumption \ref{A4:regime}, Proposition~\ref{lem2} below can be used to verify Assumption \ref{A5:kernel} for the WIRE kernel matrix. Similar lemmas for other SDR methods are provided in Supplementary Material, Appendix C.4. 
\begin{proposition}\label{lem2}
	Assume that Assumption \ref{A1:gaussian} holds and that the metric is bounded from above $m(Y, \widetilde{Y})\leq C$. Then for some positive constants $c_3, c_4, c_5$,
	\begin{equation}
		pr\{ \|\widehat{\Lambda} - \Lambda\|_{\text{max}} \geq c_3 ({{\log p}/{n}})^{1/2}\} \leq c_4 p^{-c_5-1}. 
	\end{equation}
\end{proposition}
The assumption that $m(Y, \tilde Y)\le C$ is mild as we can otherwise take a non-degenerate measurable function of $m$ (for instance $g(t) = \frac{t}{1+t}$) such that it is bounded after transformation \citep{chao2022frechet}. 
The sub-Gaussian assumption \ref{A1:gaussian} is essential for Theorem \ref{thm2} and Proposition \ref{lem2}. The convergence rate of the WIRE kernel is $O(\sqrt{\log p/n})$ under the sub-Gaussian assumption. See the proof of Proposition \ref{lem2} in the Supplementary Material.
Additionally, the sub-Gaussian assumption is commonly employed in high-dimensional statistics, as seen in the theory of LASSO \citep{wainwright2009sharp}, WIRE \citep{chao2022frechet}, and sparse regression with graphical structure \citep{yu2016sparse}.
The sub-Gaussian assumption is relatively mild, encompassing a wide range of distributions including Gaussian, Bernoulli, elliptical, and bounded types. Essentially, it applies to most distributions with light tails.
Proposition~\ref{lem2} together with Theorem~\ref{thm2} imply that 
$$\|\wh{\bm \beta}\wh{\bm \beta} ^\T - \bm \beta \bm \beta^\T\|_{\F}= O_P[\phi_d^{-1} s_0\|\Gamma_{\calS,\calS}^{-1}\|_{\infty} (\log p/n)^{1/2}],$$
which provides the rates of convergence for the central subspace estimation under the Frobenius norm. 
The convergence rate of the proposed GWIRE estimator is $O(s_0\sqrt{\log p/n})$. In the literature of SDR in the high-dimensional setting, \cite{qian2018sparse} has showed a similar rate of convergence for their sparse estimator, $O(\sqrt{s \log p/n})$, while ignoring some quantities not relating to $n$ and $p$, 
\cite{tan2020sparse,zeng2022subspace} have established the minimax rate $O(\sqrt{s \log p/n})$ for their high-dimensional SIR estimators. 
While we did not show our estimator is minimax optimal, our result is the first theoretical guarantee for high-dimensional Fr\'echet SDR problem.

\section{Synthetic and real data analysis}\label{simulation}


\subsection{Synthetic data analysis}

In this section, we investigate the numerical performance of GWIRE using synthetic data.
Recall our proposed method GWIRE uses a carefully chosen penalty to incorporate the graphical structure among predictors, 
we compare GWIRE with the following two estimates using common sparsity-inducing penalties ($\ell_1$ penalty and group-lasso penalty), 
 \begin{align}
	\widehat B_{\text{SWIRE-I}}= & \argmin_{\Theta  \in \R^{p \times p}} \frac{1}{2} \tr(\Theta^\T \widehat{\Sigma}\Theta\widehat{\Sigma})- \tr (\Theta^\T \widehat \Lambda)+\lambda \|\Theta\|_1, \label{obj:swire}\\
	\widehat B_{\text{SWIRE-II}}= & \argmin_{\Theta  \in \R^{p \times p}} \frac{1}{2} \tr(\Theta^\T \widehat{\Sigma}\Theta\widehat{\Sigma})- \tr (\Theta^\T \widehat \Lambda)+ \lambda \sum^p_{i=1}\|\Theta^{\T}e_i\|_2, \label{obj:bwire}
\end{align} 
where SWIRE is short for \textbf{S}parse \textbf{W}eighted \textbf{I}nverse \textbf{R}egression \textbf{E}nsemble. 
See more details in Appendix B.
We adopt the following criteria to assess the performance of different estimates: 
(1) General loss $\|\wh{\bm \beta}\wh{\bm \beta} ^\T - \bm \beta \bm \beta^\T\|_{\F}$, 
where ${\bm\beta}$ and $\wh{\bm \beta}$ are normalized.
(2) True recovery: $I\{S = \wh S\}$;
(3) False positive: $|\wh S \cap S^c|$; 
and (4) False negative: $|S \cap \wh S^c|$. 

We start with two examples with metric-space valued responses and graphical structured predictors.
Throughout this section, the covariates $X\in \R^p$ are sampled from a centered multivariate Gaussian distribution with covariance matrix $\Sigma$. We consider two diagonal-block covariance structures: 
(1) $\Sigma_{[25],[25]}^{(1)} = I_5 \otimes (0.16 I_5 + J_5)$, $\Sigma_{ii}^{(1)} = 1$ for $i > 25$, and $\Sigma_{ij}^{(1)} = 0$ for the other entries, where $I_5$ is a $5 \times 5$ identity matrix and $J_5$ is a $5\times 5$ matrix of
ones.
Let $\Omega^{(1)} = (\Sigma^{(1)})^{-1}$, then $\Omega^{(1)}$ is also a diagonal-block matrix with the first five blocks are 
$I_5 \otimes (0.16 I_5 + J_5)^{-1}$. 
(2) The second structure allows some deviation from Assumption \ref{A2:neighbor}. We use the $\Omega^{(2)} = \Omega^{(1)}$ from the first structure but change the entries $\Omega_{5,6}^{(2)} = \Omega_{6,5}^{(2)} = \Omega_{10,11}^{(2)} = \Omega_{11,10}^{(2)} = 0.1$, such that $\Omega^{(2)}$ connects the first block with the second block and the second block with the third one. Then let $\Sigma^{(2)} = (\Omega^{(2)})^{-1}$.

\begin{exam}\label{ex1}(Distribution)
We consider a single index model and set the coefficient
$\beta = e_1 + \cdots + e_{10}$, where $e_i$ is the $i$th canonical basis vector in $\R^p$. 
The response $Y$ is generated as the distribution with quantile function $Q_Y(\tau) = \mu_Y + \Phi^{-1} (\tau), $ where $\mu_Y\mid X \sim N(\beta^\T X, 0.1^2)$ and 
$\Phi(\cdot)$ is the cumulative distribution function of standard normal.
We use $m(Y, \widetilde{Y} )/\{1+m(Y, \widetilde{Y} )\}$ to quantify the distance between distributions $Y$ and $\widetilde Y$, 
where $m(\cdot, \cdot)$ denotes the 2-Wasserstein distance.  
\end{exam}

\begin{exam}\label{ex2}(Unit sphere)
	We consider a multiple index model with $d = 2$. 
    The coefficient vectors are $\beta_1 = 0.2(e_1 + \cdots + e_5)$ and $\beta_2 = 0.2(e_6 + \cdots + e_{10})$. The response $Y \in \R^4$ is then generated as 
    \begin{align*}
        Y_1 &= \cos(\epsilon)\sin\{\beta_1^\T (X + \bm 1_p)\} \sin\{\beta_2^\T (X + \bm 1_p)\},\\
			Y_2 &= \cos(\epsilon)\sin\{\beta_1^\T (X + \bm 1_p)\}\cos\{\beta_2^\T (X + \bm 1_p)\}, \\
			Y_3 &= \cos(\epsilon)\cos\{\beta_1^\T (X + \bm 1_p)\}, \\
			Y_4 &= \sin(\epsilon),
    \end{align*}
	where $\epsilon \sim  N(0,0.1^2)$, and $\bm 1_p$ is an all-ones vector with length $p$.
	By construction, the response $Y$ lives in a $3$-dimensional unit sphere because 
	$\|Y\|_2=1$, 
	and we use the geodesic distance $\arccos(Y^\T\widetilde Y)$ to characterize the distance between $Y$ and $\tilde Y$ in the unit sphere. 
\end{exam}

We first use the true dimension $d$ and graphical information in $\Omega$ to compare GWIRE, SWIRE-I, and SWIRE-II.
For each combination of $(n,p)$, we repeat the simulation $100$ times.
Tables~\ref{table1} and \ref{table3} present simulation results using $\Sigma^{(1)}$ for Examples~\ref{ex1} and \ref{ex2}, respectively.
The results show that 
GWIRE outperforms the other two methods in subspace estimation with the smallest general loss and variable selection with the highest true recovery rates. This is because the GWIRE approach takes into account the graphical information. 
Without considering the graphical structure among predictors, SWIRE-I tends to select more null variables yielding higher false positives, and SWIRE-II misses many active variables with higher false negatives.

\begin{table}[bt]
    \centering
	\caption{Average and standard deviation of general loss, true recovery, false positive, and false negative over $100$ simulations with various sample sizes and dimensions using the covariance matrix $\Sigma^{(1)}$ and true $d$ in Example \ref{ex1}. }
        \begin{tabular}{cccccc}
            \hline
            $(\boldsymbol{n,p})$ & {Methods} &  {General Loss} & {True Recovery} & {False Positive} & {False Negative} \\
            \hline
            $(500, 1000)$ & GWIRE & $0.09 (0.03)$ & $0.99 (0.10)$ & $0.01 (0.10)$ & $0.00 (0.00)$\\
            & SWIRE-I & $0.44 (0.09)$ & $0.71 (0.46)$ & $0.62 (1.41)$ & $0.00 (0.00)$\\
            & SWIRE-II & $0.42 (0.14)$ & $0.64 (0.48)$ & $0.20 (0.91)$ & $0.35 (0.63)$ \\
            $(500, 1500)$ & GWIRE & $0.09 (0.03)$ & $0.98 (0.14)$ & $0.03 (0.22)$ & $0.00 (0.00)$ \\
            & SWIRE-I & $0.38 (0.09)$ & $0.74 (0.44)$ & $0.47 (1.03)$ & $0.00 (0.00)$ \\
            & SWIRE-II & $0.40 (0.13)$ & $0.66 (0.48)$ & $0.19 (0.63)$ & $0.26 (0.50)$ \\
            $(500, 2000)$ & GWIRE & $0.08 (0.03)$ & $0.97 (0.17)$ & $0.03 (0.17)$ & $0.00 (0.00)$ \\
            & SWIRE-I & $0.34 (0.08)$ & $0.80 (0.40)$ & $0.32 (0.89)$ & $0.00 (0.00)$\\
            & SWIRE-II & $0.40 (0.13)$ & $0.71 (0.46)$ & $0.05 (0.26)$ & $0.27 (0.51)$ \\
            $(700, 2000)$ & GWIRE & $0.07 (0.02)$ & $0.98 (0.14)$ & $0.02 (0.14)$ & $0.00 (0.00)$ \\
            & SWIRE-I & $0.33 (0.08)$ & $0.70 (0.46)$ & $0.64 (1.48)$ & $0.00 (0.00)$ \\
            & SWIRE-II & $0.32 (0.11)$  & $0.85 (0.36)$ & $0.10 (0.46)$ & $0.10 (0.36)$ \\
            \hline
        \end{tabular}
    
	\label{table1}
\end{table}

\begin{table}
    \centering
	\caption{Average and standard deviation of general loss, true recovery, false positive, and false negative over $100$ simulations with various sample sizes and dimensions using the covariance matrix $\Sigma^{(1)}$ and true $d$ in Example \ref{ex2}. }
        \begin{tabular}{cccccc}
            \hline
            $(\boldsymbol{n,p})$ & {Methods} &  {General Loss} & {True Recovery} & {False Positive} & {False Negative} \\
            \hline
		$(500, 1000)$ & GWIRE & $0.22 (0.08)$ & $1.00 (0.00)$ & $0.00 (0.00)$ & $0.00 (0.00) $\\
		& SWIRE-I & $0.68 (0.15)$ & $0.62 (0.49)$ & $0.73 (1.65)$ & $0.09 (0.38)$\\
		& SWIRE-II & $0.51 (0.22)$ & $0.65 (0.48)$ & $0.19 (0.60)$ & $0.33 (0.67)$\\
		$(500, 1500)$ & GWIRE & $0.19 (0.06)$ & $1.00 (0.00)$ & $0.00 (0.00)$ & $0.00 (0.00) $\\
		& SWIRE-I & $0.60 (0.14)$ & $0.72 (0.45)$ & $0.84 (2.05)$ & $0.02 (0.14)$\\
		& SWIRE-II & $0.47 (0.20)$ & $0.68 (0.47)$ & $0.34 (1.17)$ & $0.21 (0.54)$\\
		$(500, 2000)$ & GWIRE & $0.17 (0.05)$ & $0.98 (0.14)$ & $0.02 (0.14)$ & $0.00 (0.00)$\\
		& SWIRE-I & $0.54 (0.14)$ & $0.78 (0.42)$ & $0.53 (1.63)$ & $0.02 (0.14)$\\
		& SWIRE-II & $0.45 (0.19)$ & $0.71 (0.46)$ & $0.19 (0.63)$ & $0.18 (0.41)$\\
		$(700, 2000)$ & GWIRE &  $0.17 (0.06)$ & $0.97 (0.17)$ & $0.06 (0.37)$ & $0.00 (0.00)$\\
		& SWIRE-I & $0.52 (0.13)$ & $0.79 (0.43)$ & $0.80 (2.85)$ & $0.01 (0.10)$\\
		& SWIRE-II & $0.38 (0.18)$ & $0.80 (0.40)$ & $0.26 (1.28)$ & $0.10 (0.33)$ \\
		\hline
		\end{tabular}
	\label{table3}
\end{table}

Recall that Assumption \ref{A2:neighbor} requires the predictor connected to the active variable should also be active. However, this assumption is difficult to check in practice. 
To assess the robustness of GWIRE when this assumption is slightly violated, 
we report the comparison results for covariates generated from $\Sigma^{(1)}$ and $\Sigma^{(2)}$ under $(n,p) = (1000,2000)$ in Table~\ref{table2}. 
The results reveal that the performance of GWIRE using $\Sigma^{(2)}$ is almost as good as using $\Sigma^{(1)}$, and GWIRE estimate has the best performance compared to SWIRE-I and SWIRE-II in all criteria. 

\begin{table}
    \centering
	\caption{Average and standard deviation of general loss, true recovery, false positive, and false negative over $100$ simulations with two covariance matrices using true $d$ when $(n,p)= (1000,2000)$ in Examples \ref{ex1} and \ref{ex2}.}%
        \begin{tabular}{ccccccc}
            \hline
            {Example} & {Structures} & {Methods} &  {General Loss} & {True Recovery} & {False Positive} & {False Negative} \\
            \hline
		\ref{ex1} & $\Sigma^{(1)}$ & GWIRE & $0.07 (0.02)$ & $0.96 (0.20)$ & $0.04 (0.20) $ & $0.00 (0.00)$\\
		&& SWIRE-I & $0.32 (0.07)$ & $0.79 (0.41)$ & $0.41 (1.11)$ & $0.00 (0.00)$ \\
		&& SWIRE-II & $0.25 (0.07)$ & $0.91 (0.29)$ & $0.10 (0.33)$ & $0.00 (0.00)$\\
		&$\Sigma^{(2)}$ & GWIRE & $0.10 (0.05)$ & $0.97 (0.17)$ & $0.03 (0.17)$ & $0.00 (0.00)$ \\
		&& SWIRE-I & $0.28 (0.06)$ & $0.75 (0.44)$ & $0.87 (2.35)$ & $0.00 (0.00)$\\
		&& SWIRE-II & $0.24 (0.07)$ & $0.89 (0.31)$ & $0.24 (0.81)$ & $0.00 (0.00)$\\
		\ref{ex2} & $\Sigma^{(1)}$ & GWIRE & $0.17 (0.07) $ & $1.00 (0.00)$ & $0.00 (0.00)$ & $0.00 (0.00)$\\
		&& SWIRE-I & $0.51 (0.11)$ & $0.75 (0.44)$ & $0.76 (1.97)$ & $0.00 (0.00)$\\
		&& SWIRE-II & $0.33 (0.14)$ & $0.83 (0.38)$ & $0.29 (1.17)$ & $0.07 (0.29)$\\
		&$\Sigma^{(2)}$ & GWIRE & $0.22 (0.14)$ & $0.98 (0.16)$ & $0.01 (0.10)$ & $0.05 (0.50)$ \\
		&& SWIRE-I & $0.48 (0.11)$ & $0.69 (0.46)$ & $0.80 (1.94)$ & $0.01 (0.10)$\\
		&& SWIRE-II & $0.30 (0.13)$ & $0.81 (0.39)$ & $0.49 (1.55)$ & $0.05 (0.26)$\\
		\hline
		\end{tabular}
	\label{table2}
\end{table}

\begin{table}
	\centering
	\caption{Proportions of correct structural dimension $d$ over $100$ simulations in Examples~\ref{ex1} and ~\ref{ex2}.}
	\begin{tabular}{ccccc}
			\hline
			 & \multicolumn{2}{c}{Example~\ref{ex1} ($\boldsymbol{d=1}$)}  & \multicolumn{2}{c}{Example~\ref{ex2} ($\boldsymbol{d=2}$)}\\
			\cline{2-5}
			 ($\boldsymbol{n,p}$) & $\boldsymbol{\Sigma^{(1)}}$ & $\boldsymbol{\Sigma^{(2)}}$  & $\boldsymbol{\Sigma^{(1)}}$ & $\boldsymbol{\Sigma^{(2)}}$\\
			 \hline
			$(500, 1000)$ & $0.92$ & $0.76$ &
			$0.92$ & $0.84$\\
			$(500, 1500)$ & $0.71$ & $0.54$ & 
			$0.88$ & $0.76$\\
			$(500, 2000)$ & $0.71$ & $0.64$ & 
			$0.70$ & $0.71$ \\
			$(700, 2000)$ & $1.00$ & $1.00$ & 
			$0.94$ & $0.94$\\
			$(1000,2000)$ & $1.00$ & $1.00$ & 
			$1.00$ & $0.98$\\
			\hline
			\end{tabular}	
	\label{table5}
\end{table}

When the structural dimension $d$ is unknown, we use $\hat{d}$ in Equation \eqref{eq:d-hat} to estimate $d$ for GWIRE.
Table~\ref{table5} shows proportions of correct estimation of $d$ 
over 100 simulations. When the sample size increases with fixed $p=2000$, the percentages of correct estimation of $d$ achieve as high as $100\%$. 
When the graphical information among predictors is unknown,
we use the graphical Lasso 
\citep{friedman2008sparse} to estimate the precision matrix and obtain the neighborhood information from $\widehat\Omega$.
Table~\ref{table7} presents the estimation results of Examples \ref{ex1} and \ref{ex2} using estimated graph information.
It shows that with the estimated graph information, GWIRE still has good performance in terms of estimation and variable selection with small general losses and high recovery rates. 

\begin{table}
    \centering
	\caption{Average and standard deviation of general loss and true recovery using true $d$ for GWIRE over $100$ simulations with the estimated precision matrix as neighbor information when $(n,p)= (1000,2000)$ in Examples \ref{ex1} and \ref{ex2}. }
		\begin{tabular}{cccccc}
			\hline
			&&\multicolumn{2}{c}{Example \ref{ex1}}& \multicolumn{2}{c}{Example \ref{ex2}}\\
			\cline{3-6}
			($\boldsymbol{n,p}$) & {Neighbor} &  {General Loss} & {True Recovery} & {General Loss} & {True Recovery} \\
			\hline
			$(500, 1000)$ & $\wh{\Omega}^{(1)}$ &  $0.08 (0.03)$ & $0.99 (0.10)$ & $0.22 (0.08)$ & $0.97 (0.17)$\\
			& $\wh{\Omega}^{(2)}$ &  $0.09 (0.03)$ & $0.97 (0.17)$ &$0.25 (0.10)$ & $0.96 (0.20)$\\
			$(500, 1500)$ & $\wh{\Omega}^{(1)}$ &  $0.08 (0.03)$ & $1.00 (0.00)$ & $0.18 (0.06)$ & $0.98 (0.14)$\\
			& $\wh{\Omega}^{(2)}$ &  $0.08 (0.03)$ & $0.98 (0.14)$ & $0.22 (0.11)$ & $0.96 (0.20)$ \\
			$(500, 2000)$ & $\wh{\Omega}^{(1)}$ &  $0.08 (0.03)$ & $0.96 (0.20)$ & $0.17 (0.06)$ & $0.99 (0.10)$\\
			& $\wh{\Omega}^{(1)}$ &  $0.08 (0.03)$ & $0.99 (0.10)$  & $0.17 (0.05)$ & $0.98 (0.14)$\\
			$(700, 2000)$ & $\wh{\Omega}^{(1)}$ &  $0.07 (0.02)$ & $0.98 (0.14)$ & $0.17 (0.06)$ & $0.99 (0.10)$ \\
		 & $\wh{\Omega}^{(2)}$ &  $0.07 (0.02)$ & $0.98 (0.14)$ & $0.18 (0.07)$ & $0.96 (0.20)$ \\
			$(1000, 2000)$ & $\wh{\Omega}^{(1)}$ &  $0.07 (0.02)$ & $0.99 (0.10)$ & $0.17 (0.07)$ & $1.00 (0.00)$ \\
			& $\wh{\Omega}^{(2)}$ & $0.07 (0.02)$ & $0.99 (0.10)$ & $0.17 (0.08)$ & $0.99 (0.10)$\\
			\hline
			\end{tabular}
	\label{table7}
\end{table}

We further consider two examples that have univariate responses and graph-structured predictors. 
Our goal is to illustrate the effectiveness of the proposed optimization \eqref{obj2} with classical SDR kernel matrices (SIR and CUME), whose estimators are referred to as GSIR and GCUME.

\begin{exam}\label{ex3}
$Y = \exp(\beta^\T X + 0.5\epsilon)$, where  $\beta = e_1 + \cdots + e_{10}$, $X\sim N(0, \Sigma^{(1)})$,
$\epsilon \sim N(0,1)$, and $X\perp \epsilon$. 
\end{exam}

\begin{exam}\label{ex4}
$Y = \exp(\beta_1^\T X)\text{sign}(\beta_2^\T X) + 0.2\epsilon$, where $\beta_1 = e_1 + \cdots + e_{5}$ and $\beta_2 = e_6 + \cdots + e_{10}$, $X\sim N(0, \Sigma^{(1)})$, $\epsilon \sim N(0,1)$, and $X\perp \epsilon$. 
\end{exam}

 \begin{table}
    \centering
	\def~{\hphantom{0}}
	\caption{Average and standard deviation of general loss and true recovery using true $d$ over $100$ simulations with various sample sizes and dimensions with the covariance matrix $\Sigma^{(1)}$ in Examples \ref{ex3} and \ref{ex4}. We use 10 slices for GSIR.}
		\begin{tabular}{cccccc}
			\hline
			&&\multicolumn{2}{c}{Example \ref{ex3}}& \multicolumn{2}{c}{Example \ref{ex4}}\\
			\cline{3-6}
			($\boldsymbol{n,p}$) & {Methods} &  {General Loss} & {True Recovery} & {General Loss} & {True Recovery} \\
			\hline
		$(500, 1000)$ & 
		GSIR & $0.05 (0.02)$ & $1.00 (0.00)$ & $0.17 (0.18)$ & $0.96 (0.20)$\\
		& GCUME & $0.08 (0.02)$ & $1.00 (0.00)$ & $0.34 (0.07)$ & $0.91 (0.29)$\\
		$(500, 1500)$ & 
		GSIR & $0.05 (0.02)$ & $1.00 (0.00)$ & $0.12 (0.04)$ & $0.99 (0.10)$\\
		& GCUME & $0.07 (0.02)$ & $0.99 (0.10)$ & $0.20 (0.06)$ & $0.92 (0.27)$ \\
		$(500, 2000)$ & 
		GSIR & $0.05 (0.02)$ & $0.99 (0.10)$ & $0.12 (0.03)$ & $0.99 (0.10)$\\
		& GCUME & $0.06 (0.01)$ & $1.00 (0.00)$ & $0.18 (0.08)$ & $0.91 (0.29)$\\
		$(700, 2000)$ & 
		GSIR & $0.04 (0.01)$ & $0.99 (0.10)$ & $0.13 (0.13)$ & $0.98 (0.14)$ \\
		& GCUME & $0.06 (0.02)$ & $1.00 (0.00)$  & $0.17 (0.06)$ & $0.93 (0.26)$ \\
		$(1000, 2000)$ & 
		GSIR & $0.03 (0.01)$ & $1.00 (0.00)$& $0.13 (0.14)$ & $0.99 (0.10)$ \\
		& GCUME & $0.06 (0.01)$ & $1.00 (0.00)$ & $0.17 (0.05)$ & $0.95 (0.22)$ \\
		\hline
	\end{tabular}
	\label{table6}
\end{table}

Table~\ref{table6} displays the estimation results of GSIR and GCUME under different $(n,p)$. 
GSIR and GCUME have excellent performance in terms of small general loss and high recovery rate.

\subsection{Real data analysis: bike rental data}
The real data analysis aims to illustrate the performance of the proposed method in real applications. Our proposal is the first method to consider high-dimensional predictors in Fr\'echet SDR.
We apply our proposed GWIRE approach to analyze daily bike rental data \citep{fanaee2014event} at Washington, D.C. \footnote{UCI Machine Learning Repository: \url{https://archive.ics.uci.edu/ml/datasets/bike+sharing+dataset}}. 
This dataset spans the years 2011 and 2012 for 730 days after deleting one day with missing values. 
The response vector for each day is the 24 observed frequency of bike rental counts. 
The 24-hour rental counts can be viewed as a discrete distribution, a random object in a metric space. Hence, the proposed method is particularly well-suited for this application.
Each response characterizes a categorical distribution taking values in $\{0, ..., 23\}$. 
For two discrete distributions $P = (p_0, \cdots, p_{23})$ and $Q = (q_0, \cdots, q_{23})$, using the Hellinger distance provides a more straightforward measure of their distance. The Hellinger distance is defined as $H(P, Q) = \left\{\sum^{23}_{i=0} \left(\sqrt{p_i} - \sqrt{q_i}\right)^2/2\right\}^{1/2}$. We normalize it by using $H(P, Q)/\{1+H(P, Q)\}$ to measure their distance. If these distributions are considered as continuous distributions, we could also use the 2-Wasserstein distance, but that would require quantile functions of the discrete distributions. Hence, we opt to use the Hellinger distance to measure the distance between two discrete distributions.

We consider the following eight predictors from the data: 
\begin{itemize}
	\item Holiday: Indicator of a public holiday celebrated in Washington D.C.
	\item Working: Indicator of neither the weekend nor a holiday
	\item Temp: Daily mean temperature
	\item Atemp: Feeling temperature
	\item Humid: Daily mean humidity
	\item Wind: Daily mean windspeed
	\item BW: Indicator of bad weather (misty and/or cloudy)
	\item RBW: Indicator of really bad weather (snowy and/or rainy)
\end{itemize}
According to the data description, the predictors Holiday and Working are correlated because a holiday day is also a non-working day. 
The other six predictors characterizing the weather in different aspects are also corrected. 
For instance, an indicator of bad weather often means low temperature and high-speed wind. 
Hence, it's reasonable to assume a graphical structure among these eight predictors, and we estimate the neighborhood information using graphical Lasso \citep{friedman2008sparse}.

Before applying our proposal GWIRE to this dataset, we first standardize the predictors by removing the mean and scaling to unit variance. 
To illustrate the performance of GWIRE, we create additional $1000$ independent predictors from the standard normal distribution. 
That is, we have in total of $1008$ predictors, of which at most eight of them are relevant to the response. 
We apply GWIRE and its competitors SWIRE-I and SWIRE-II to estimate the matrix $B$. 
We consider the structural dimension to be 2 and extract two directions $\hat{\beta}_1$ and $\hat{\beta}_2$ for visualization and illustration. 
The three methods (GWIRE, SWIRE-I, SWIRE-II) do not select Humid, Wind, and any noise variables, and Table~\ref{table:realdata} presents the corresponding coefficients for the original 8 predictors.
The independent normal noise predictors are unrelated to the response. The proposed method can effectively identify meaningful predictors while ignoring the noise predictors. This demonstration precisely showcases the performance of the proposed method in high-dimensional data. Since it does not select any noise predictors, the validity of the real data analysis remains unaffected.
The GWIRE leverages graphical structure and selects six features except for Humid and Wind, while SWIRE-I and SWIRE-II only select four and two features, respectively. 
This is because when predictors are highly correlated, GWIRE tends to choose a group of correlated variables, while SIWRE and SWIRE-II only pick some predictors from the correlated predictors. 
Besides the difference in variable selection, the three estimates provide similar information from extracted directions. 
The first direction has high loading on the predictor Working, while the second has high loading on predictors characterizing the temperature, especially Temp and Atemp. 
Hence, we interpret the sufficient predictors as the "working" and "temperature" indices.

\begin{figure}
	\centering
	{\includegraphics[width=0.5\textwidth]{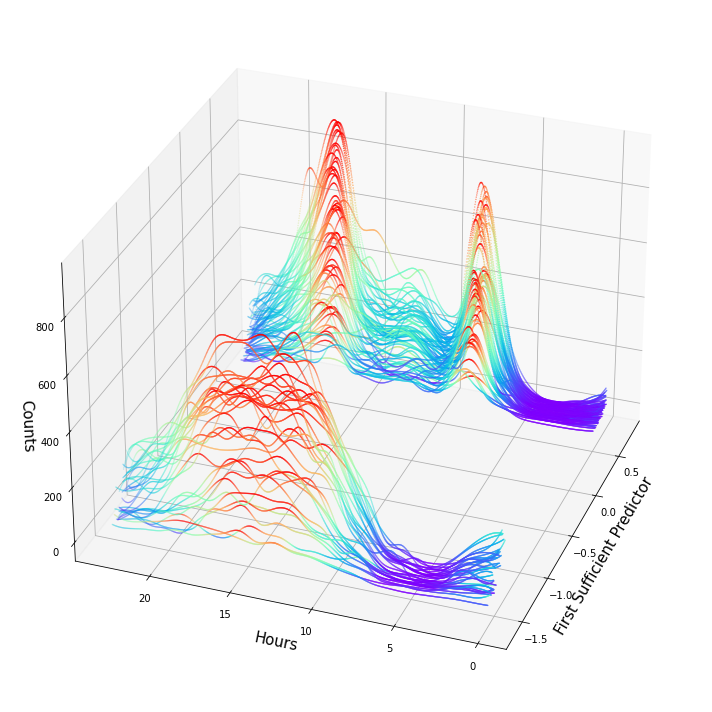} }
	\caption{Plot of rental bike counts versus the first sufficient predictor in randomly selected 100 days.}
	\label{fig:realdata1}
\end{figure}

\begin{figure}
	\centering
	\subfigure[Working Days]{{\includegraphics[width=0.4\textwidth]{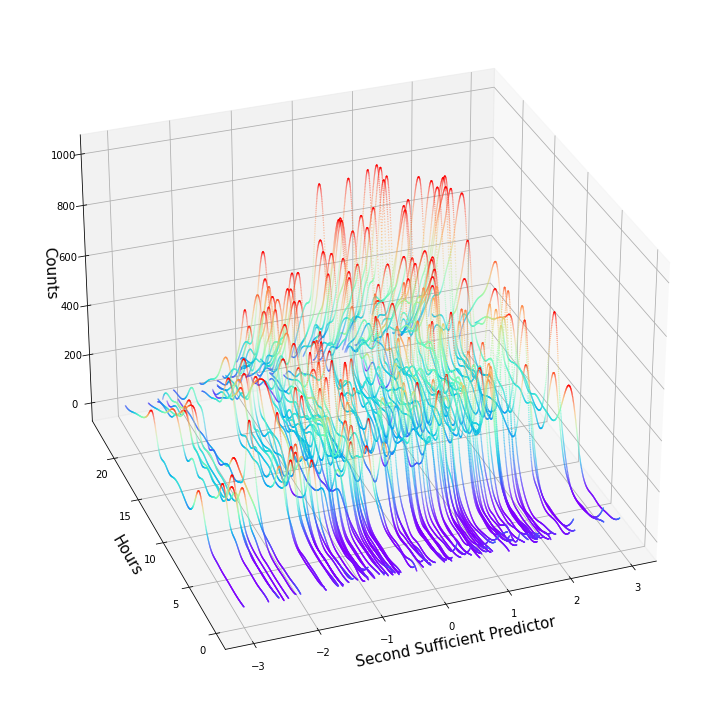} }}%
	\subfigure[Non-Working Days]{{\includegraphics[width=0.4\textwidth]{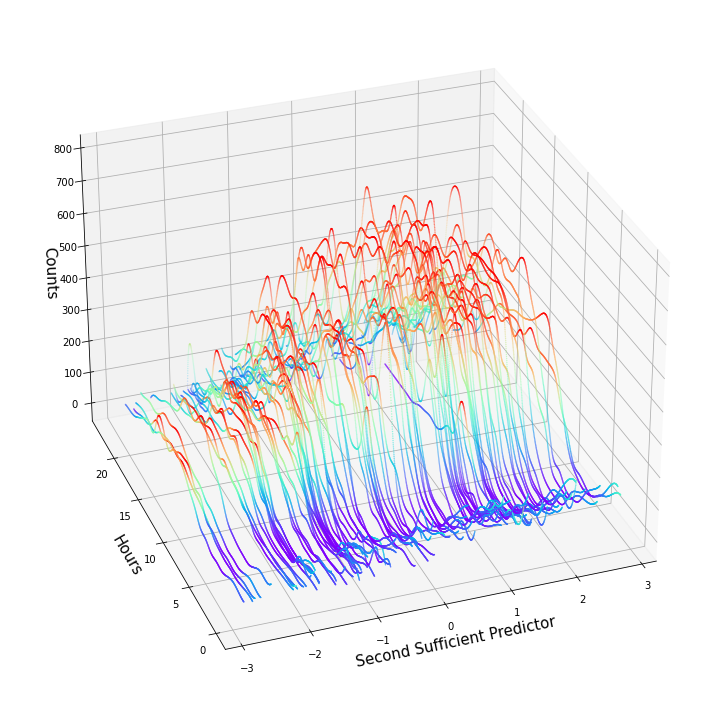} }}\\
	\subfigure[Working Days]{{\includegraphics[width=0.4\textwidth]{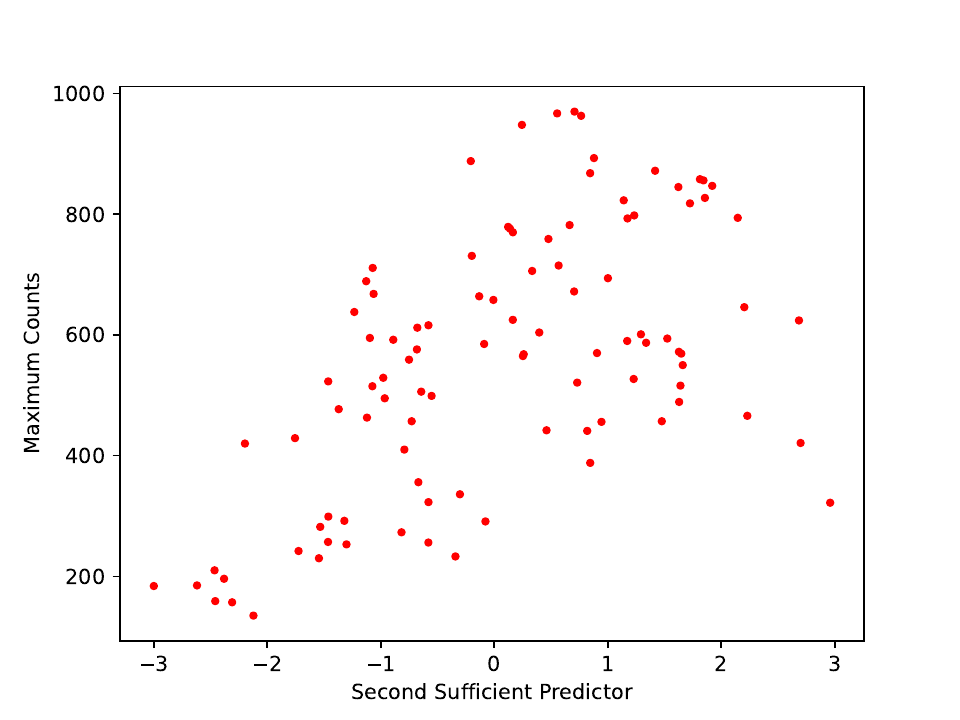} }}%
	\subfigure[Non-Working Days]{{\includegraphics[width=0.4\textwidth]{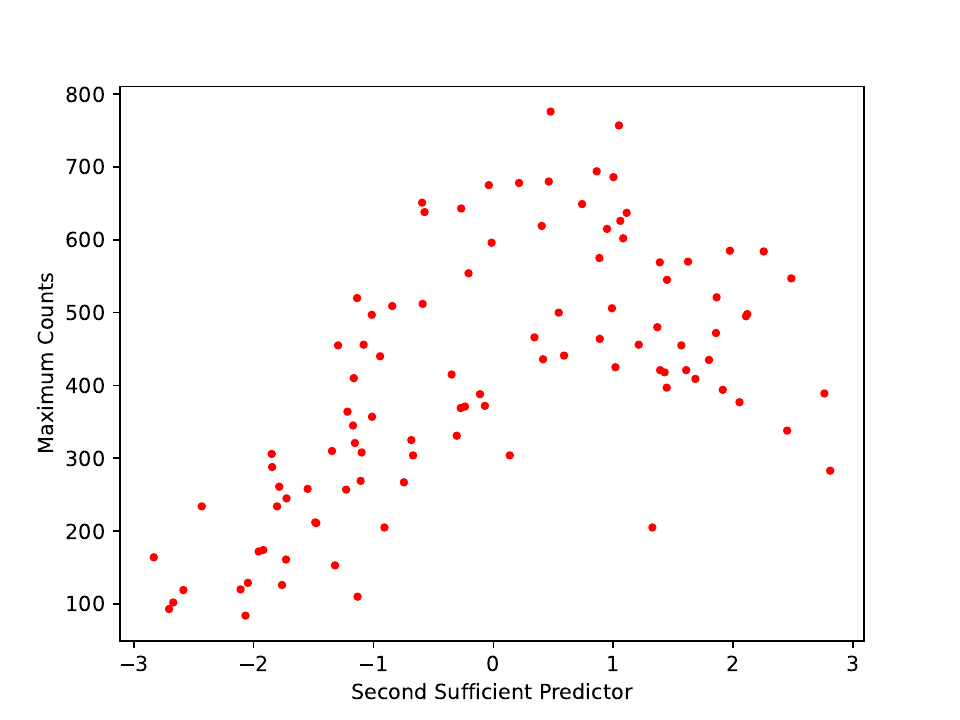} }}
	\caption{ 
	(a-b) Plots of rental bike counts versus the second sufficient predictor for working days and non-working days based on the GWIRE estimates; (c-d) Plots of the maximum counts of bike rental distributions versus the second sufficient predictor on working days and non-working days. 
	}
	\label{fig:realdata2}
\end{figure}

We further plot the bike rental counts versus the sufficient predictors $X^{\T}\hat{\beta}_1$ and $X^{\T}\hat{\beta}_2$ from GWIRE.
Figure~\ref{fig:realdata1} shows that the first sufficient predictor captures the shape of curves, which reflects different rental bike patterns on working and non-working days. 
Figure~\ref{fig:realdata2} (a, b) show that with the increase of the second sufficient predictor, the peak of curves first goes up and then goes down beyond a certain point. 
This reflects that people are more willing to use bicycles when the temperature is appropriate and less likely to use bicycles when it is too cold or too hot. 
This phenomenon is made more clear in Figure~\ref{fig:realdata2} (c, d).

\begin{table}
	\centering
	\caption{Coefficients of the first two directions for three methods: GWIRE, SWIRE-I, and SWIRE-II.
	}
		\begin{tabular}{lcccccc}
			\hline
			& \multicolumn{2}{c}{GWIRE} & \multicolumn{2}{c}{SWIRE-I} & \multicolumn{2}{c}{SWIRE-II}\\
			\cline{2-7}
					&  $\boldsymbol{\hat{\beta}_1}$ & $\boldsymbol{\hat{\beta}_2}$ &
					   $\boldsymbol{\hat{\beta}_1}$ & $\boldsymbol{\hat{\beta}_2}$ &
					   $\boldsymbol{\hat{\beta}_1}$ & $\boldsymbol{\hat{\beta}_2}$ \\
					   \hline
					   Holiday & 0.02 & -0.02 & 0.02 & 0.00  & 
					   0.00 & 0.00\\
					   Working &  1.00 & -0.05 & 1.00 & 0.00 & 
					   1.00  & -0.00\\
					   Temp & 0.03 & 0.75 & 
					   0.00 & 0.93 & 
					   0.00 & 1.00\\
					   Atemp & 0.04 & 0.65 & 
					   0.00 & 0.37 & 
					   0.00 & 0.00\\
					   Humid & 0.00 & 0.00 & 
					   0.00 & 0.00 & 
					   0.00 & 0.00\\
					   Wind & 0.00 & 0.00 & 0.00 & 0.00 & 0.00 & 0.00\\
					   BW & -0.02 & -0.11 & 
					   0.00 & 0.00 & 
					   0.00 & 0.00\\
					   RBW & -0.00 & -0.02 & 0.00 & 0.00 & 0.00 & 0.00\\
					   \hline
		   \end{tabular}
	\label{table:realdata}
\end{table}

\section{Discussion}
This paper constructs a general framework for \frechet SDR with a metric-space valued response and high-dimensional predictors. 
In particular, we propose a novel optimization problem to avoid the inverse of a large covariance matrix and leverage graphical information among predictors. 
We also establish the subspace estimation and variable selection consistency for the proposed estimator. 
Furthermore, most SDR approaches including SIR and CUME can be used in our procedure.  
Additional to group penalty, we can consider the minimax concave penalty to mitigate estimation bias \citep{weng2024sparse}. 
We conclude with two open directions associated with this work. 
(a) It is of independent interest to construct confidence intervals (ellipsoids) and tests for coefficients of SDR.
(b) Another direction is to extend our procedure to situations where predictors are tensors, functions, or random objects in metric space. 
\bibliographystyle{plainnat}
\bibliography{wire}
\end{document}